%% file: conference_101719.tex
\documentclass[10pt, conference]{IEEEtran}
\IEEEoverridecommandlockouts
\usepackage{amsmath,amssymb,amsfonts}
\usepackage{tabularx}
\usepackage{algorithmic}
\usepackage{graphicx}
\usepackage{textcomp}
\usepackage{xcolor}
\usepackage{bm}
\usepackage{caption}
\usepackage{subcaption}

\newcommand{\const}{\bm{\Omega}^M}
\newcommand{\cint}{\sZ_\sC}

\input{math_commands.tex}

\IEEEoverridecommandlockouts\IEEEpubid{\makebox[\columnwidth]{ 978-1-6654-3540-6/22~\copyright~2022 IEEE \hfill} \hspace{\columnsep}\makebox[\columnwidth]{ }}
    
\begin{document}

\title{Learning Perturbations for Soft-Output Linear MIMO Demappers

\thanks{*Work completed at Qualcomm AI Research.}
\thanks{Qualcomm AI Research is an initiative of Qualcomm Technologies, Inc.}}

\author{\IEEEauthorblockN{Daniel E. Worrall}
\IEEEauthorblockA{
\textit{Qualcomm AI Research*}\\
Amsterdam, The Netherlands \\
dworrall@qti.qualcomm.com}
\and
\IEEEauthorblockN{Markus Peschl}
\IEEEauthorblockA{\textit{Qualcomm AI Research}\\
Amsterdam, The Netherlands \\
mpeschl@qti.qualcomm.com}
\and
\IEEEauthorblockN{Arash Behboodi}
\IEEEauthorblockA{\textit{Qualcomm AI Research}\\
Amsterdam, The Netherlands \\
behboodi@qti.qualcomm.com}
\and
\IEEEauthorblockN{Roberto Bondesan}
\IEEEauthorblockA{\textit{Qualcomm AI Research}\\
Amsterdam, The Netherlands \\
rbondesa@qti.qualcomm.com}
}

\maketitle

\begin{abstract}
Tree-based demappers for multiple-input multiple-output (MIMO) detection such as the sphere decoder can achieve near-optimal performance but incur high computational cost due to their sequential nature.
In this paper, we propose the perturbed linear demapper (PLM), which is a novel data-driven model for computing soft outputs in parallel. To achieve this, the PLM learns a distribution centered on an initial linear estimate and a log-likelihood ratio clipping parameter using end-to-end Bayesian optimization. Furthermore, we show that lattice-reduction can be naturally incorporated into the PLM pipeline, which allows to trade off computational cost against coded block error rate reduction. We find that the optimized PLM can achieve near maximum-likelihood (ML) performance in Rayleigh channels, making it an efficient alternative to tree-based demappers.

\end{abstract}




\section{Introduction}
Multiple-input and multiple-output detection is a fundamental problem in modern  wireless communication enabling high data rates for transmission of multiple streams over multiple antennas \cite{MIMO2003}. MIMO demappers are typically required to calculate soft probability estimates of transmitted bits, which can then be used as informative priors for \emph{error correction codes}, such as LDPC codes. This, however, comes at the cost of increased computational complexity \cite{Studer2008}, which is why it is crucial to develop fixed, low complexity detectors \cite{Barbero2006} that can achieve near-maximum-likelihood (near-ML) performance in terms of coded bit error rate (BER).



The simplest of such low complexity detectors are \emph{linear MIMO detectors}, based on the Zero-Forcing (ZF) or Minimum-Mean Square Error (MMSE) paradigm. Linear MIMO detectors fail to account for inter-stream interference and for this reason are not typically considered a viable candidate for obtaining soft outputs in high correlation channels. To tackle this, tree based methods iteratively search the solution space for a set of likely candidates and have proven to attain near-ML performance at the cost of added complexity \cite{Studer2010, Qi2012ParallelHT}. Nonetheless, the search procedure of these methods relies on manually designed heuristics, which offers little flexibility in terms of end-to-end optimization with respect to different channel statistics and error correction procedures.
Recently, there has been an increased interest in learning-based methods for MIMO detection, which have been shown to achieve state of the art results for predicting hard-output solutions  \cite{Liao2020, pratik2020re}. 
On the other hand, the design of learnable soft-output MIMO demappers with low complexity poses a much larger challenge and has yet to be explored. 

To circumvent this issue of complexity, we propose the \textit{perturbed linear demapper}, which learns a parametric distribution centered on a lattice reduced hard-output solution to generate a diverse set of candidates. Previously, lattice reduction (LR) aided linear demappers \cite{Wuebben2004} have been shown to improve soft MIMO detection by perturbing an initial solution \cite{Ponnampalam2007} or employing Schnorr-Euchner enumeration to obtain a set of most likely transmitted symbols \cite{Hyunsub2014}. Our key insight is that this idea can be generalized using minimal added complexity to allow for more flexibility as well as generate more reliable log-likelihood ratios (LLRs) for subsequent error correction. We achieve this by jointly learning parameters for (i) LLR clipping and (ii) the shape of the perturbation distribution using Bayesian optimization (BO) \cite{snoek2012practical} on simulated data. 


\textbf{Contributions:} We take a soft-output LR-aided linear detector with LLR clipping on the output and use Bayesian optimization to select an optimal perturbation distribution and LLR clipping parameters with respect to the bit error rate, post-LDPC decoding. We show that our method achieves comparable BER to the low-complexity tree-based Parallel Fixed-Complexity Sphere Decoder (PFSD) \cite{Qi2012ParallelHT}, while adding the benefit
of a candidate generation that is fully parallelizable and adaptable to channel statistics.


\section{Background}
\textit{Notation}: Matrices are boldface uppercase, vectors are boldface lowercase, and scalars are unboldened. Superscript $H$ denotes the conjugate transpose. The 2-norm of $\rvx$ is $\|\rvx\|$. $\rmI_{M}$ is the $M\times M$ identity matrix and $\bm{0}$ is a vector of zeros and $\1$ a vector of ones. $x_{j,b}$ denotes the $b$\textsuperscript{th} bit in the graymap decoding of $x_j$, the $j$\textsuperscript{th} dimension of $\rvx$. $\sZ_\sC := \sZ + i\sZ$ is the set of complex integers, where $i=\sqrt{-1}$ is the imaginary unit, $\E$ is the expectation operator, and $\dot\cup$ denotes the disjoint union operator.

\subsection{MIMO Demapping}
We consider MIMO systems based on the famous V-BLAST architecture \cite{Wolniansky1998}. This is a multiple antenna system with $M$ transmit and $N\geq M$ receive antennas. A coded bit-stream is mapped to transmit symbol vectors $\rvx\in\bm{\Omega}^M$, where $\bm{\Omega} \subset \alpha \cint - \frac{\alpha}{2}(\1+i\1)$ is a $2^Q$-QAM complex constellation for $\alpha > 0$ and even $Q$, the number of bits per symbol $x_j$. $\alpha$ is chosen such that QAM symbols are normalized to unit power per dimension (i.e.\ $\E[\rvx\rvx^H]=\rmI_{M}$). Each symbol vector $\rvx$ has a corresponding bit-decoding in $\{-1,1\}^{M\times Q}$, where $x_{j,b}$ is the $b$\textsuperscript{th} bit in the $j$\textsuperscript{th} dimension/layer of $\rvx$. We assume a flat-fading environment with a block fading channel (constant gains per frame), changing independently frame-to-frame. We therefore model a single slot of the time-discrete complex baseband model as a linear system with additive white complex Gaussian noise (AWGN):
\begin{align}
    \rvy = \rmH \rvx + \rvn. \label{eq:channel-model}
\end{align}
$\rmH$ is an $N\times M$ matrix, representing a Rayleigh fading channel, with i.i.d. complex Gaussian fading gain entries $H_{ij} \sim \gC\gN(0, 1/\sigma^2)$ for $\sigma > 0$ and $\rvn$ is a zero-mean circular-symmetric complex Gaussian noise vector with identity covariance, (i.e.\ $\E[\rvn\rvn^H]=\rmI_{N}$). This channel model has per-stream received SNR proportional to $ \log (1 / \sigma^2)$. We also assume perfect channel estimation at the receiver. While some works lift the $M \times N$ complex system to a $2M \times 2N$ real system, we instead choose to perform all calculations in complex space.

In soft-output MIMO detection, we aim to compute bitwise log-likelihood ratios (LLRs)
\begin{align}
    L(x_{j,b}) = \log \frac{p(x_{j,b}=1 | \rvy, \rmH)}{p(x_{j,b}=-1 | \rvy, \rmH)}, \label{eq:llr}
\end{align}
which can be fed as soft-information into a downstream error correction algorithm. This contrasts the hard-output problem, known as \emph{integer least squares} \cite{Hassibi2005}, where we solve
\begin{align}
    \rvx^* = \argmin_{\rvx \in \const} \Vert \rvy - \rmH\rvx \Vert^2.
\end{align}
Given our AWGN model, the LLRs can be rewritten
\begin{align}
    L(x_{j,b}) &= \lambda_{\gX_{j,b}^+} - \lambda_{\gX_{j,b}^-} \label{eq:llr-expanded} \\
    \lambda_{\gX_{j,b}^\pm} &= \log \sum_{\rvx\in\gX_{j,b}^\pm} \exp \left \{ -\Vert \rvy - \rmH\rvx \Vert^2 \right \} \label{eq:lse}
\end{align}
where $\gX_{j,b}^+ \subset \bm{\Omega}^M$ is the subset of points $\rvx$ with bit $x_{j,b}=1$ and likewise for $\gX_{j,b}^-$ and $x_{j,b}=-1$. The corresponding bit of the larger of the two log likelihoods (in absolute value) is referred to as the \emph{hypothesis bit} and the smaller, the \emph{counter-hypothesis bit}. A large part of the complexity of soft-output demappers is in computing counterhypothesis bits, since while a single hypothesis vector $\rvx^*$ contains all the hypothesis bits, the counterhypothesis bits may be scattered across up to $MQ$ counterhypothesis vectors. Sometimes, \Eqref{eq:lse} is replaced with the \emph{max-log approximation}, which approximates the sum by the highest mass point as
\begin{align}
    \lambda_{\gX_{j,b}^\pm} &\simeq \min_{\rvx\in\gX_{j,b}^\pm} \left \Vert \rvy - \rmH\rvx \right \Vert^2 \label{eq:mla}.
\end{align}
This approximation, a variant of the hard-output problem, was shown in \cite{Studer2008} to only incur a small performance loss of about 0.25 dB. Even with this simplification, however, computing the solution to \Eqref{eq:mla} is itself an NP-hard problem.

\subsection{Linear Receivers}
Linear detectors demap the received vector $\rvy$ in two steps: left-multiplying \Eqref{eq:channel-model} by a linear filter matrix $\rmG$, followed by quantization to the constellation. Mathematically
\begin{align}
    \label{eq:babai}
    \rvx_{\text{linear}} = \lfloor \rmG \rvy \rceil_{\const}
\end{align}
where $\lfloor \bullet \rceil_{\const}$ denotes quantization and remapping into the constellation $\const$. Different linear detectors make different choices about $\rmG$. For a finite set $\gX$ we define the quantization operator for a real scalar variable $x$ as
\begin{align}
    \lfloor x \rceil_{\gX} = \min_{y \in \gX} |x - y|.
\end{align}
For complex scalars, we quantize the real and imaginary components independently and for vectors, we quantize each dimension independently too.

\subsubsection{Zero-Forcing Detector}
The Zero-Forcing detector uses the Moore-Penrose pseudo-inverse of the channel matrix:
\begin{align}
    \rmG_{\text{ZF}} = \rmH^+ = (\rmH^H\rmH)^{-1}\rmH^H .
\end{align}

\subsubsection{MMSE Detector}
The MMSE detector is a more sophisticated linear receiver, which takes into account of the noise statistics of the channel. This sets
\begin{align}
    \rmG_{\text{MMSE}} = (\rmH^H\rmH + \rmI)^{-1}\rmH^H.
\end{align}
Note that we normalize the channel to use unit variance noise in our model, so $\sigma^2$ is implicitly defined in $\rmH$. We can also write the detector as a Zero-Forcing detector over the \emph{extended channel}, where
\begin{align}
    \underline{\rmG_{\text{MMSE}}} = \underline{\rmH}^+, \quad \underline{\rmH} = \begin{bmatrix} 
    \rmH \\
    \rmI_{M\times M}
    \end{bmatrix}, \quad \underline{\rvy} = \begin{bmatrix} 
    \rvy \\
    \bm{0}_{M}
    \end{bmatrix},
\end{align}
satisfying the identity
\begin{align}
    \rmG_{\text{MMSE}} \rvy = \underline{\rmG_{\text{MMSE}}} \, \underline{\rvy}.
\end{align}

\begin{figure*}[t!]
    \centering
    \includegraphics[width=0.88\linewidth]{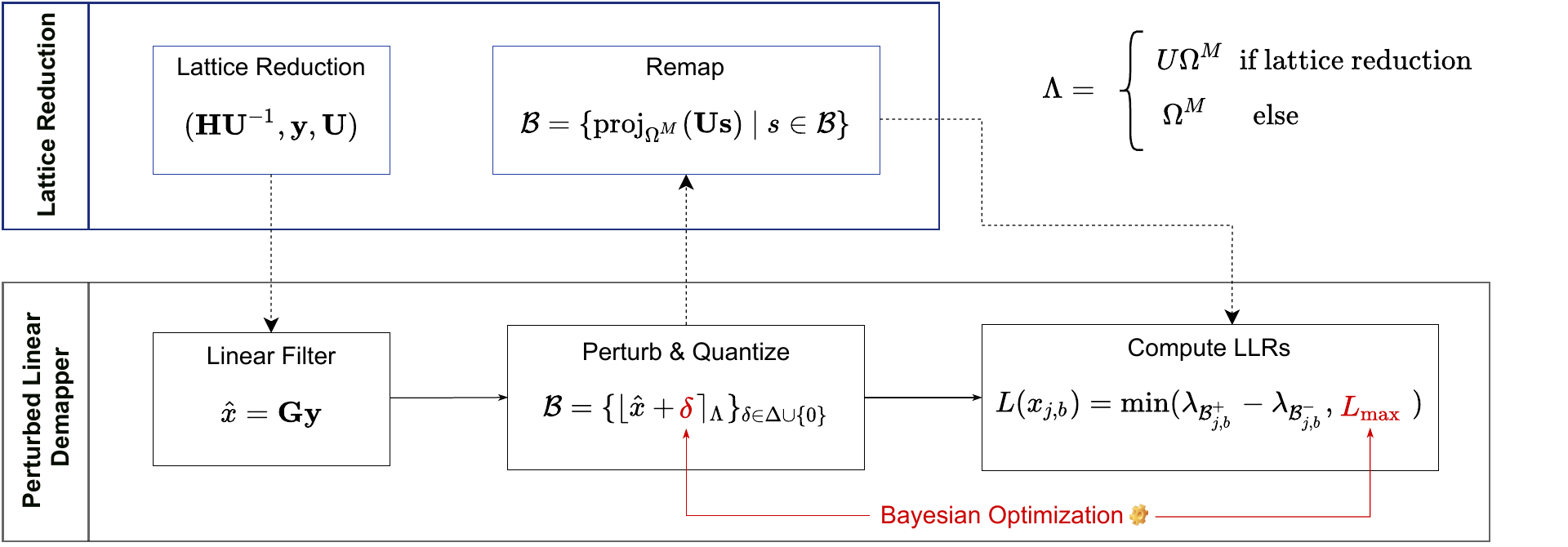}
    \caption{Schematic of the PLM with (optional) lattice reduction. Candidate generation is performed by applying perturbations to a linear initial solution, from which LLRs are computed using the max-log approximation. Bayesian optimization is applied to jointly learn LLR clipping parameters and perturbation distributions in an end-to-end fashion.}
    \label{fig:plmvis}
\end{figure*}

\subsubsection{Soft-output ZF/MMSE Detector}
The above linear receivers only produce hard outputs. We can compute soft-outputs by rewriting the channel model as \cite{StuderSoft}
\begin{align}
    \rvy = x_j\rvh_j + \left(\sum_{i\neq j} x_i \rvh_i + \rvn\right) = x_j\rvh_j + \rvn_j,
\end{align}
for $\rvh_1, \rvh_2, ...$ the columns of $\rmH$ and $\rvn_j$ a new noise vector treating layers $i\neq j$ as interference. If we marginalize out the symbols $x_{i\neq j}$, we are left with a new channel for $p(\rvy | x_j \rvh_j)$ with covariance $\rmK = \rmI + \sum_{i\neq j} \rvh_i\rvh_i^H$. We can turn this into a one dimensional problem by left-multiplying by a per-channel decorrelating filter $\rvw_j$:
\begin{align}
    \rvw_j^H \rvy = x_j\rvw_j^H \rvh_j  + \rvw_j^H\rvn_j, \quad \rvw_j = \rmK^{-1} \rvh_j
\end{align}
and approximating $p(\rvw_j^H\rvy | x_j \rvw_j^H\rvh_j)$ with a 1D complex Gaussian with matched variance. We then have
\begin{align}
    \lambda_{\gX_{j,b}^\pm} \simeq \log \sum_{x_j \in \bm{\Omega}_{b}^\pm} \exp \left \{ -(\rvw_j^H \rvy - x_j\rvw_j^H \rvh_j)^2 /\sigma_j^2 \right \}
\end{align}
where $\sigma_j^2 = \rvh_j^H \rmK^{-1} \rvh_j$ is the variance $\rmK$ projected on to $\rvw$ and $\bm{\Omega}_{b}^\pm$ is the subset of constellation points with bit $b$ set to $\pm 1$. If instead of using $(\rmH, \rvy)$ in the above we used $(\underline{\rmH}, \underline{\rvy})$, then we would have a soft-output MMSE detector.

\subsection{Lattice Reduction} \label{sec:lattice-reduction}
The columns of $\rmH$ form the \emph{basis} of a \emph{lattice} 
\begin{align}
    \gL = \left \{\sum_{j=1}^M x_j \rvh_j \middle | x_j \in \sZ_\sC,\, j=1,...,M \right \}.
\end{align}
The lattice basis is non-unique, and we can transform between bases by right-multiplication of $\rmH$ with a \emph{unimodular matrix} 
\begin{align}
    \rmU \in GL(M, \sZ_\sC).
\end{align}
where $GL(M, \sZ_\sC)$, the $\emph{general linear group}$ over the complex integers, is the set of all matrices with complex integer entries and unit determinant. Each element of $GL(M, \sZ_\sC)$, thus each unimodular matrix, has an inverse, which is also unimodular. Due to the unit determinant, right-multiplication by $\rmU$ is volume-preserving. 

The unimodular matrices provide a change of basis for the lattice $\gL$ and thus a reparametrization of the channel $\rmH$ as $\rmH \mapsto \rmH\rmU^{-1}$. Crucially, it turns out that most MIMO detectors\footnote{A counterexample is a maximum likelihood detector, such as the sphere decoder.} are not invariant under this reparameterization, indicating that certain choices of $\rmU$ are better than others. Concretely, if the columns of $\rmH$ are orthogonal, there is no inter-stream interference and linear detectors are optimal. It has therefore been shown \cite{Wuebben2004} that a good choice of $\rmU$ is one that approximately orthogonalizes the columns of $\rmH$, measured in terms of the \emph{log orthogonality defect}
\begin{align}
    \log \delta(\rmH) = \sum_{j=1}^M \log \Vert \rvh_j \| - \frac{1}{2}\log \det \rmH^H \rmH,
\end{align}
a non-negative quantity, reaching zero when the columns of $\rmH$ are orthogonal. Detectors using the lattice reduced channel matrix $\rmH\rmU^{-1}$ are called \emph{LR-aided} \cite{Wuebben2004}. We can apply LR to linear detectors as shown in figure \ref{fig:plmvis}. The only differences are that we use a different channel $\rmH\rmU^{-1}$ and quantize to the \emph{reduced lattice} $\rmU\gL$ before transforming back and bounding to the original constellation $\const$.
If we let $\bm{\beta} = \frac{\alpha}{2}(\1+i\1)$, we can efficiently compute the lattice quantization operator of a lattice $\rmU\gL$ as \cite{Ponnampalam2007}
\begin{align}
    \lfloor \rvz \rceil_{\rmU\gL} = \alpha \lfloor \alpha^{-1} (\rvz - \rmU\bm{\beta}) \rceil_{\cint^m} + \rmU\bm{\beta}.
\end{align}

\begin{figure*}[!t]
    \centering
    \includegraphics[width=0.65\linewidth]{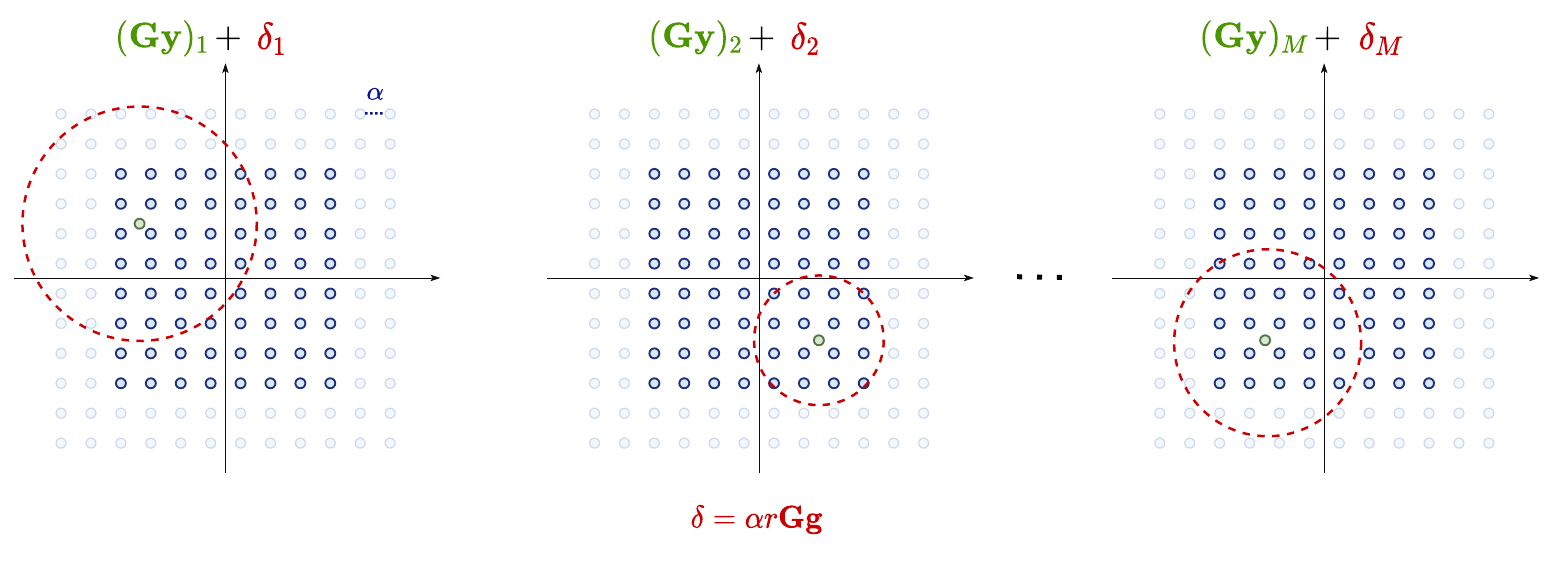}
    \caption{Visualization of the PLM in complex space. The lattice corresponding to the channel is visualized in blue, where only the points that lie inside the QAM constellation are opaque. The PLM obtains a linear estimate (green) and then samples from a perturbation distribution (red). In the case of Gaussian perturbations, our method corresponds to sampling from a circular distribution with different variances for each detection layer.}
    \label{fig:testvis}
\end{figure*}

\section{Perturbed Linear Detection}

\subsection{Motivation}
The complexity of the log-sum-exp computation in \Eqref{eq:lse} is determined by the size of the search-space $|\gX_{j,b}^\pm|=|\bm{\Omega}^M|/2 = 2^{QM-1}$, scaling exponentially in $M$ and $Q$. In practice, to reduce this complexity we approximate it by replacing each set $\gX_{j,b}^\pm$ with a smaller \emph{candidate set} $\gB_{j,b}^\pm$. LLR computation can then be broken down into two steps:
\begin{enumerate}
    \item Candidate set $\gB_{j,b}^\pm$ generation 
    \item Compute LLR $\lambda_{\gB_{j,b}^\pm}$: \Eqref{eq:lse} or \Eqref{eq:mla}.
\end{enumerate}

For the candidate set generation, we propose a new class of model, which we call a \emph{perturbed linear model} (PLM). For the LLR computation, we leverage classical methods, simply plugging our candidate sets $\gB_{j,b}^\pm$ into \Eqref{eq:lse} or \Eqref{eq:mla}.

\subsection{Perturbed Linear Model}
The PLM generates a candidate set $\gB$ by taking the output $\rmG\rvy$ of a linear receiver, adding small perturbations $\bm{\delta}$ to it, and then quantizes back on to the constellation
\begin{align}
    \gB = \left \{ \lfloor \rmG \rvy + \bm{\delta} \rceil_{\const} \right \}_{\bm{\delta} \in \bm{\Delta} \cup \{\bm{0}\}}.
\end{align}
Using samples from the set $\bm{\Delta} \cup \bm{0}$ instead of $\bm{\Delta}$ ensures that we always include the Babai-estimate of \eqref{eq:babai}. We can then sift this set $\gB$ into individual candidate sets for each bit $j,b$ set to $\pm 1$ as
\begin{align}
    \gB_{j,b}^\pm = \{\rvx \in \gB \mid x_{j,b} = \pm 1
    \}. 
\end{align}
Of course, $\gB_{j,b}^+ \,\dot\cup\, \gB_{j,b}^- =\gB$ for all $j$ and $b$. The motivation for this is that the hypothesis $\rvx^*$ lies close to the linear receiver output $\rmG\rvy$, and can be found within a ball centered on $\rmG\rvy$ spanned by $\bm{\Delta}$. Finding the counter-hypotheses, however, is non-trivial but we can search for them in a region around the hypothesis. Thus how we choose to populate $\bm{\Delta}$ is important.

 We consider three different choices for perturbation generation. Firstly, we sample in an ellipsoid ($\bm{\delta}_e$), given by:
 \begin{align}
     \bm{\delta}_e &= r \rmG (u^{-2M\kappa} \rvr)  && u\sim \gU(0,1), \rvr \sim \gC\gS_M \label{eq:ellipsoid}
 \end{align}
 where $r > 0$ is a radius parameter, $u$ is a real uniform random variable in interval $[0,1]$, $\kappa > 0$ is a radial weighting factor and $\rvr$ is a random vector drawn from $\gC\gS_M$, the uniform distribution on a complex sphere of dimension $M$. The parameters $r$ and $\kappa$ are learnable. \Eqref{eq:ellipsoid} draws samples from a ball distribution and passes them through $\rmG$, which makes the distribution of samples ellipsoidal. When $\kappa > 1$, the density of samples close to 0 increases. We correlate dimensions of samples with $\rmG$ since, say, for Zero-Forcing detectors with $\rm\rmH^+$, decoded symbol vectors have the form $\rvx = \rmG \rvy - \rmG \rvn$, which is centered on $\rmG\rvy$ with additive noise $\rmG \rvn$.
 Second, we consider sampling a non-isotropic Gaussian ($\bm{\delta}_g$) defined as:
 \begin{align}
     \bm{\delta}_g &= r\rmG \rvg &&\rvg \sim \gC\gN(0,1^2), \label{eq:gaussian} 
 \end{align}
 where  $\rvg$ is a complex Gaussian vector from a zero-centered unit-variance circular symmetric complex density. \Eqref{eq:gaussian} draws standard complex Gaussian samples and passes them through $\rmG$ as well. Similarly, $r$ is a learnable parameter.

 Finally, we consider determinstic perturbations in the form of a $K$-QAM-constellation in each dimension ($\bm{\delta}_s$):
 \begin{align}
     \bm{\delta}_s &= q\rve_i &&  i \in \{1,...,M\}, q\in \bm{\Omega}_K \label{eq:ponnampolam}
 \end{align}
 where $\rve_i$ is a one-hot vector with a one in position $i$, and $\bm{\Omega}_K$ is a one dimensional $K$-QAM constellation. \Eqref{eq:ponnampolam} generates candidates by drawing a square around $\rmG \rvy$ in each dimension, which is a similar setup to \cite{Ponnampalam2007}. However, \cite{Ponnampalam2007} does not study larger QAM-constellation sizes and thus expands only 4 points in each dimension. A key benefit of the perturbative method is that the generation of perturbations can be parallelized, leading to fast implementation.

\subsection{LLR computation}
After subset generation, we compute LLRs using the max-log approximation \cite{Hochwald2003} as 
\begin{align}
\lambda_{\gB_{j,b}^\pm} &= \min_{\rvx\in\gB_{j,b}^\pm} \Vert \rvy - \rmH\rvx \Vert^2 .
\end{align}
In some cases, one of the sets $\gB_{j,b}^+$ or $\gB_{j,b}^-$ is empty and the corresponding $\lambda$ defaults to $\infty$. We found we could improve our LLR outputs, when used for downstream LDPC decoding with LLR clipping to a default value $L_{\text{max}}$, such that
\begin{align}
L(x_{j,b}) = \min (\lambda_{\gB_{j,b}^+} - \lambda_{\gB_{j,b}^-}, L_{\text{max}}).
\end{align}
We optimize $L_\text{max}$ for each SNR, storing it in a lookup table.

\subsection{Bayesian Optimization}

The whole demapper from input bits to output bits after LDPC decoding can be treated as a black box depending on a few parameters. These are, namely, $L_\text{max}$ and the perturbation distribution parameters $r$ and $\kappa$. We denote all these parameters $\bm{\theta} = \{r, \kappa, L_\text{max}\}$. We optimize these, using Bayesian optimization (BO) \cite{Osborne09gaussianprocesses}, a machine learning-based global optimization technique. In particular, for each SNR we measure the coded BER $L_\text{BER}$ averaged over a subset of $B$ instances of the channel in \Eqref{eq:channel-model} for a given value of $\bm{\theta}$. 
Note that here we use BO as an optimizer for the learning problem of finding parameters $\bm{\theta}$ that give small $L_{\text{BER}}$ when estimated during deployment over a test set of channel instances.
In more details, 
we fit a Gaussian process \cite{rasmussen2005} with Mat\'ern covariance kernel to this training data of input--output pairs $\{(\bm{\theta}_k, L_{\text{BER},k})\}_k$, which provides a differentiable Bayesian regression estimator of the $\bm{\theta} \mapsto L_{\text{BER},k}$ mapping. This can then be used to choose the next value of $\bm{\theta}$ in the optimization process by maximizing an acquisition function. Our implementation used the openly available \textsc{Scikit-optimize} library\footnote{\texttt{https://github.com/scikit-optimize/scikit-optimize}}.

The BO process is run as a separate `offline' step to find the optimal values (up to noise and optimization constraints) of $\bm{\theta}$ for every SNR. At deployment time, we simply look up each $\bm{\theta}$ given the SNR of the channel, which in this work we assume to be known. Also, our approach does not require differentiability of the pipeline and is far more efficient than running a grid-search, which becomes computationally infeasible in problems of even just a handful of variables.

\subsection{Efficient Lattice Reduction}
The PLM does not require LR to operate, but we found that including an LR step to preprocess the channel helps. In particular it is important to identify a good hypothesis, around which the perturbation distribution is defined. In the LR-aided setup the channel $\rmH$ is replaced with $\rmH\rmU^{-1}$ and we change quantization accordingly, see Section \ref{sec:lattice-reduction}.

LR itself is an NP-hard problem and thus we opt for an efficient approximate algorithm to compute $\rmU$. We use the method in Algorithm 5 of \cite{Cong2013}, which is low-, fixed-complexity and parallelizable. Empirically, we find the required number of Gram-Schmidt iterations to be $\lceil \sqrt{M} \rceil$. Note that we only need to compute $\rmU$ once for each $\rmH$ and in a practical setting the cost of LR will be amortized across multiple channel uses, where $\rmH$ is assumed to be constant; although, in our setup we opt of a worst-case scenario and recompute $\rmH$ per transmission.

\subsection{Complexity Analysis}
 We present a complexity outline of the proposed approach in terms of floating point operations (FLOPS) in table \ref{tab:complexity}. For comparison, we follow the same conventions as \cite{Zu2012} and assume that 
 \begin{itemize}
     \item A complex $m \times n$  times $n \times p$ matrix multiplication takes $8mnp$ FLOPS,
     \item A QR decomposition of a $m\times n$ matrix takes $16(n^2m-nm^2+\frac{1}{3}m^3)$ FLOPS,
     \item LR takes roughly $1.6$ times as many FLOPS as a single QR decomposition.
     \item Random sampling from a Gaussian or uniform distribution takes up to constant $c$ FLOPS.
 \end{itemize} Furthermore, we provide example FLOP counts for $M=N=4$ and $|\Omega| = 256$ with $|\mathcal{B}|=M\cdot 256$ perturbations in total. As can be seen, LR, MMSE detection, and perturbation generations make up for a relatively small amount of the overall cost. While quantization, remapping and Euclidean distance calculations are more expensive, we note that they scale linearly with $|\mathcal{B}|$. Furthermore, as we will show in section \ref{subsec:lrexp}, the PLM yields reasonable performance without lattice reduction. The PLM therefore opens two avenues for a complexity-performance trade off, where LR offers a constant complexity trade off with gains depending on the orthogonality defect of the channel and reducing the size of $|\mathcal{B}|$ provides a linear complexity reduction. Aside from this flexibility, the PLM is fully parallelizable which makes it a promising candidate for efficient large-scale MIMO detection. As opposed to tree-based demappers, this is because the list generation in the PLM does not rely on any interference cancellation, but instead on independent samples from a fixed probability distribution.


\begin{table}[b]
    \centering
    {\begin{tabularx}{\columnwidth}{|X||c|X|}
    \hline
    Operation & FLOPS & Example\\
    \hline \hline
        Lattice Reduction & $\!\begin{aligned}[t]
    &25.6 (M^2N+MN^2+(1/3)N^2)
    \end{aligned}$     & 15497\\ \hline
        
        MMSE &      $\!\begin{aligned}[t]
    &2(M+N)^3 - 2(M+N)^2  \\&+16M(M+N)^2 \\ &+ 8M(M+N) + M + N
    \end{aligned}$        
        & 5256\\ \hline
        
        Perturbations & $2c|\mathcal{B}|M$ & 8192 \\ \hline
        Quantize & $10|\mathcal{B}|M+8M^2$ & 41088 \\ \hline
        Remap & $8|\mathcal{B}|M^2$ & 131072 \\ \hline
        Calculate Euclidean Distances & 
         $\!\begin{aligned}[t]
        &16(MN^2+M^2N+(1/3)M^3) \\ &+8MN+|\mathcal{B}|M(4M^2+2M)
        \end{aligned}$ 
         & 76246 \\ \hline
        LLR Computation & $|\mathcal{B}|\log |\bm{\Omega}|M$ & 32768\\ \hline
    \end{tabularx}}
        \caption{Complexity of the perturbed linear demapper.}
    \label{tab:complexity}

\end{table}

\section{Experiments}

\subsection{Coded channel}
\begin{figure*}[!htb]
    \centering
    \begin{subfigure}[b]{0.45\textwidth}
        \centering
        \includegraphics[width=0.77\textwidth]{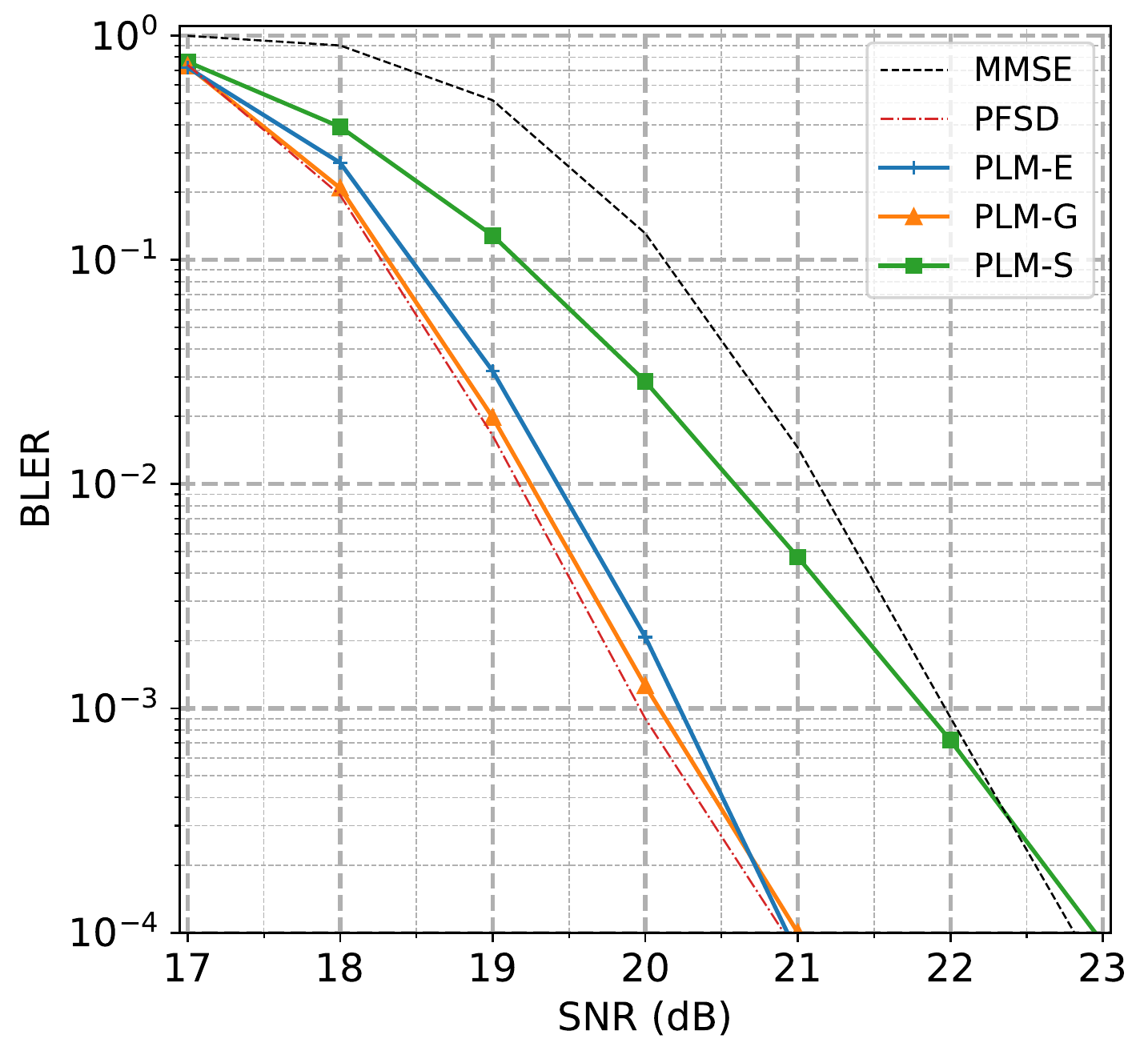}
        \caption{We show PFSD \cite{Qi2012ParallelHT}, MMSE, \& PLM with \textbf{S}quare, \textbf{E}llipsoidal, and \textbf{G}aussian perturbations.}
        \label{fig:coded-channel}
    \end{subfigure}
    \hfill
    \begin{subfigure}[b]{0.45\textwidth}
        \centering
        \includegraphics[width=0.77\textwidth]{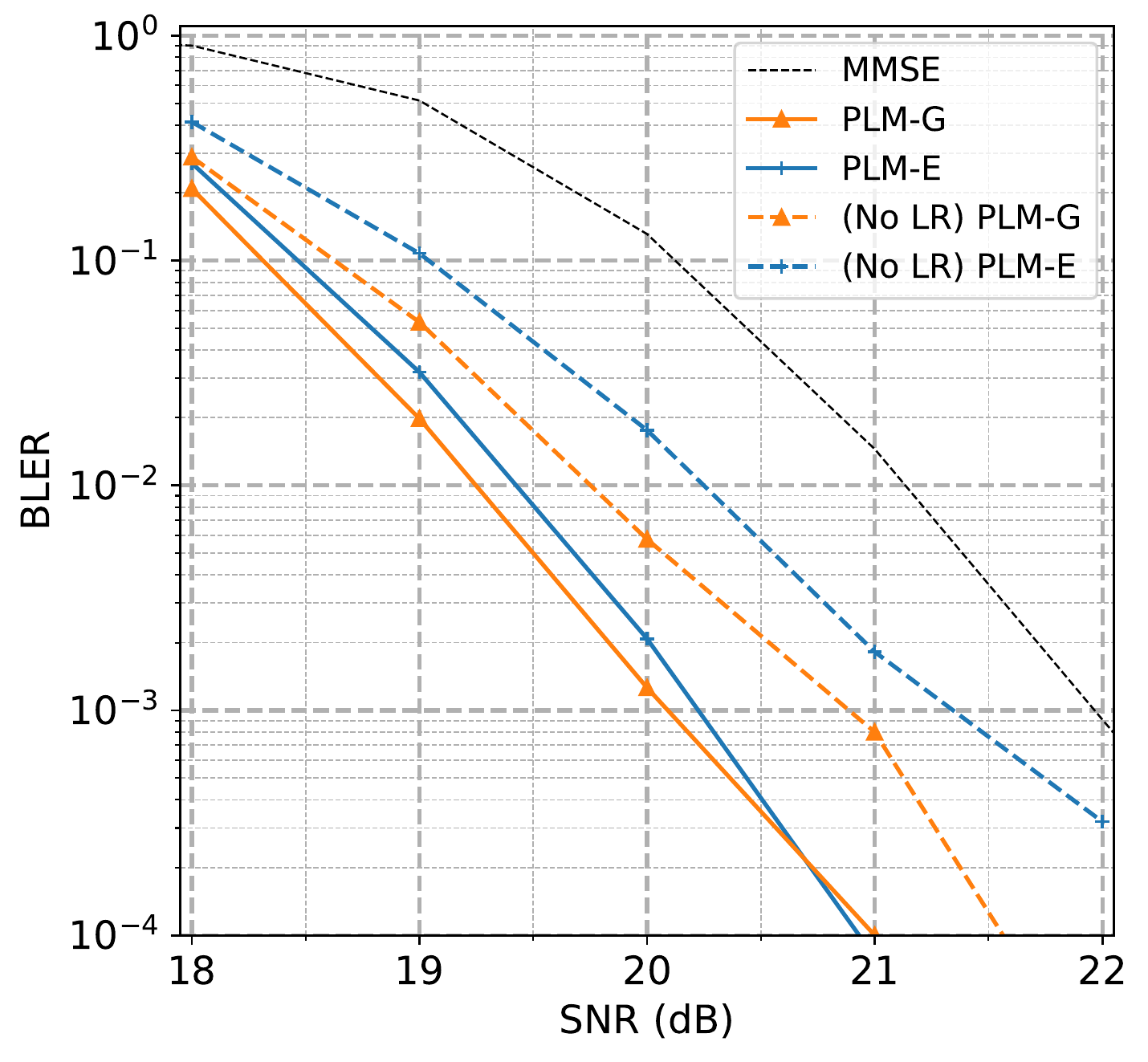}
        \caption{PLM model comparison for ellipsoidal and Gaussian perturbations with and without LR. We see that LR improves BLER.}
        \label{fig:lr-comparison}
    \end{subfigure}
    \caption{Performance comparison of the PLM in a $4\times 4$, $256$-QAM MIMO problem. Coded block error rates are shown for different SNR (dB) using LDPC error correction of rate 1/3 and base graph number $1$.}
    \label{fig:coded-channel-experiments}
\end{figure*}

We measure block error rate (BLER) for a $4\times 4$ 256-QAM  system with rate 1/3 LDPC coded Rayleigh fading channel and a block size of 704 for a variety of models. Our models are a baseline MMSE receiver, PFSD \cite{Qi2012ParallelHT}, and our LR-aided PLM under different perturbation schemes. We matched the number of list candidates per layer (256) in all models and illustrate the results in Figure \ref{fig:coded-channel}. As can be seen, the PLM with Gaussian perturbations performs on par with PFSD, closely trailed by the ellipsoidal perturbations. On the other hand square perturbations perform worse and approach the MMSE curve for higher SNR. We conjecture that this is the case due to the difficulty of finding representative counterhypotheses in large QAM constellations for each bit. While square perturbations guarantee to sample a fixed subset close to the relatively likely initial seed solution, they fail to obtain counterhypothesis bit solutions and as a result provide worse uncertainty estimates to the LDPC decoder.

\subsection{Efficacy of lattice reduction}
\label{subsec:lrexp}
When comparing our lattice-reduced PLM against its non lattice-reduced counterpart Figure \ref{fig:lr-comparison}, we see that LR helps in terms of BLER. This performance boost comes merely from the difference in quantization regions induced by LR. Aside from this, the learned perturbations with and without LR are similar, aside from their different basis representation. While this performance gain of 0.5-1.0 dB is significant, we note however, that the PLM without LR still performs surprisingly well and at the same time yields only half of the complexity due to the omission of the remapping step. 


\section{Conclusion}
In this work, we have proposed the perturbed linear demapper (PLM), a data-driven model for soft-output MIMO detection. In order to efficiently generate a list of candidates for the LLR computation, the model makes use of parametric distributions for perturbing the output solution of a simple linear demapper. We show that we can leverage Bayesian optimization to learn optimal perturbation and postprocessing parameters to achieve near-ML performance in terms of coded BLER. 
Our method achieves comparable BLER to PFSD \cite{Qi2012ParallelHT}, but adds the benefit of a fully parallelizable candidate generation as well as offers adaptability to channel statistics. Furthermore, the PLM can utilize lattice-reduction to strictly improve its performance at some added complexity cost.

It is a question for future work to find if there are better perturbation distributions, which are performant in terms of BLER and also in terms of sample efficiency when we reduce the number of candidates in the set $\gB$. In particular, we consider the parametrization of perturbations using neural networks as an especially promising future direction.

\bibliographystyle{IEEEtran}
\bibliography{IEEEabrv,bibliography}

\end{document}

%% file: math_commands.tex
\usepackage{amsmath,amsfonts,bm}

\def\eqref#1{equation~\ref{#1}}
\def\Eqref#1{Equation~\ref{#1}}

\def\1{\bm{1}}

\def\rve{{\mathbf{e}}}

\def\rvg{{\mathbf{g}}}
\def\rvh{{\mathbf{h}}}

\def\rvn{{\mathbf{n}}}

\def\rvr{{\mathbf{r}}}

\def\rvw{{\mathbf{w}}}
\def\rvx{{\mathbf{x}}}
\def\rvy{{\mathbf{y}}}
\def\rvz{{\mathbf{z}}}

\def\rmG{{\mathbf{G}}}
\def\rmH{{\mathbf{H}}}
\def\rmI{{\mathbf{I}}}

\def\rmK{{\mathbf{K}}}

\def\rmU{{\mathbf{U}}}

\DeclareMathAlphabet{\mathsfit}{\encodingdefault}{\sfdefault}{m}{sl}
\SetMathAlphabet{\mathsfit}{bold}{\encodingdefault}{\sfdefault}{bx}{n}

\def\gB{{\mathcal{B}}}
\def\gC{{\mathcal{C}}}

\def\gL{{\mathcal{L}}}

\def\gN{{\mathcal{N}}}

\def\gS{{\mathcal{S}}}

\def\gU{{\mathcal{U}}}

\def\gX{{\mathcal{X}}}

\def\sC{{\mathbb{C}}}

\def\sZ{{\mathbb{Z}}}

\usepackage{todonotes}

\usepackage{amsthm}
\usepackage{amssymb}

\theoremstyle{definition}

\theoremstyle{remark}

\def\E{{\mathbb{E}}}

\DeclareMathOperator*{\argmin}{arg\,min}